\input{aipcheck}
 
\documentclass[
    ,final            
  ]
  {aipproc}

\layoutstyle{6x9}

\begin{document}

\title{The EMC Effect and Short-Range Correlations}


\author{Misak M. Sargsian}{
  address={Florida International University, Miami, FL 33199}
}

\begin{abstract}
We overview the progress made  in studies of EMC  and short range correlation~(SRC) effects with the  special emphasis 
given to the recent observation of the correlation between the slope of the EMC ratio at Bjorken $x<1$ 
and the scale factor of the same ratio at $x>1$ that measures the strength of the SRCs in nuclei. 
This correlation may  indicate the larger modification of nucleons with  higher  momentum  thus making  the  
nucleon virtuality as the  most relevant parameter of medium modifications.
To check this conjecture we study the implication of several properties of high momentum component of 
the nuclear wave function on the characteristics of EMC effect.  We observe two main reasons for 
the EMC-SRC correlation: first, the decrease of the contribution from the nuclear mean field due to the 
increase, with A,  the fraction of the high momentum  component of nuclear wave function. 
Second,  the  increase of the  medium modification of nucleons in SRC.  
Our {\em main prediction} however is the increase of the   proton contribution  to the EMC effect for large A asymmetric nuclei.
This prediction is based on the recent observation of the  strong dominance of $pn$ SRCs 
 in the high momentum component of nuclear wave function. Our preliminary calculation based on this prediction of 
 the excess of  energetic and modified protons in large A nuclei describes reasonably well the  main features of the 
 observed  EMC-SRC correlation.
 
 \end{abstract}

\maketitle



{\bf The EMC Effect:}
The discovery of the nuclear EMC effect\cite{EMC1} is one of the unique  cases in which the 
explicit degrees of QCD  is intertwined with the picture of 
nucleus consisting of  hadrons.  However the specific dynamics of the modification of 
partonic distributions~(PDFs)  in bound nucleon  have  not yet been identified with certainty. 
The quantity under consideration  is the ratio of inclusive cross sections of nuclei A and deuteron 
measured in deep inelastic kinematics  at $x<1$  
(corrected by  the factor, $f$  which accounts  for the unequal  numbers  of protons and neutrons):
\vspace{-0.2cm}
\begin{equation}
R_{EMC}(x,Q^2)  =  {2\cdot \sigma_{eA}\over A\cdot \sigma_{ed}}f(x,Q^2).
\vspace{-0.2cm}
\end{equation}
Since $R_{EMC}$ is measuring the ratio of nucleon PDFs in A and the deuteron it was expected that 
$R_{EMC} =1$.  However the first experiments\cite{EMC1,EMC3} found it substantially less than unity in the region of 
$0.3 < x < 0.8$.  Later experiments\cite{EMC5} were able 
to quantify the magnitude of the EMC effect as proportional to $A$ or to the average nuclear density 
defined as $\rho(A) = 3A/4\pi R_e^3$, with $R_e^2 = 5\langle r^2\rangle/3$, where 
$\langle r^2\rangle$ is the nuclear RMS radius measured in elastic $eA$ scattering.  
The recent measurements of  EMC effects at Jefferson Lab reached to unprecedented accuracy\cite{EMC7}. 
These measurements demonstrated that the early observation of the proportionality of the EMC effect to 
average nuclear density is not valid, with $^9Be$ nucleus clearly out of sync with other nuclei. However the 
simple monotonic $A$ dependence agreed well with the all measured nuclei. Thus these measurements demonstrated 
that for the EMC effect the $A$ and average density dependences are not equivalent.  Another result of 
the new experiment was the observation of no $Q^2$ dependence of the depletion of $R_{EMC}$ for 
$4 < Q^2 < 6$~GeV$^2$.  
The parallel theoretical development in EMC studies was the realization that 
due to the charge Z the nucleus  has a Coulomb field  which 
in the reference frame in which the nucleus  has a large momentum is transformed into the
field of equivalent photons. If $Z\alpha_{em}$ is not small then the equivalent photons carry 
finite fraction of nuclear momentum.  The account of this momentum fraction in heavy 
nuclei as compared to the deuteron removed some of the EMC effect  for medium to heavy nuclei 
at $0.3<x<0.5$ region.  Thus the genuine medium modification is  associated with the 
depletion  of the nuclear structure function only in  $0.5 < x < 0.8$ region\cite{Frankfurt:2010cb}.

Summarizing, the latest progress in EMC studies  indicates 
that;  (a) the size of the effect is proportional to A but not to the average nuclear density;   (b) the account for the  
Coulomb  effects narrows the range of the EMC effect to $05 < x < 0.8$, the region corresponding to 
the scattering off the  bound nucleon with large initial momenta;  ({c})  no apparent $Q^2$ dependence is 
observed for $4 < Q^2 < 6$~GeV$^2$ which may provide important constraint on the potential 
mechanism of  EMC effects.

{\bf Short Range Nucleon Correlations in Nuclei:}  SRCs are considered one of the 
 most elusive features of the ground state nuclear wave functions.  It is expected not to be 
 probed directly with any low  energy probe. However advent of the high energy probes 
 allowed a significant progress in isolating and studying  the dynamical nature of 
 2N SRCs (for recent reviews see \cite{srcrev,srcprogress}).   One of the methods in probing 
 2N SRCs  is studying  high $Q^2$ inclusive $A(e,e')X$ scattering at $x>1.4$ in which case virtual photon 
 scatters off the bound nucleon with momenta exceeding $k_{F}(A)$ \cite{ms01,hnm}.  
 If the scattering indeed happens with the nucleon from  2N SRC then the 
 prediction is that the ratio of the inclusive cross sections of nucleus A  and the  deuteron should exhibit a 
 plateau\cite{FS8188,FSDS}.  Such a plateau was observed in both  SLAC\cite{FSDS} and recent 
 JLab\cite{Kim,Fomin} measurements.  Another recent news from SRC studies is the observation of 
 a strong (by factor of 20)  dominance of $pn$ relative to $pp$ and $nn$ SRC's in the range of 
 the  bound nucleon momenta $k_F < p < 600$~MeV/c\cite{isosrc,EIP4}. This observation was an indication 
 that at the distances relevant to the above momentum range the NN force is dominated by tensor interaction.
 This gave a new meaning to the above mentioned ratios: 
\vspace{-0.2cm}
 \begin{equation}
 a_2(A) = {2\cdot \sigma_{eA}\over A\cdot \sigma_{ed}},
\vspace{-0.2cm}
 \label{a2}
 \end{equation}
 which  now represent  (up to the SRC center of mass motion  effect) the probability of finding 2N SRCs in the nucleus A. The 
 observed strong disbalance of $pn$ and $pp$/$nn$ SRCs allowed also to suggest new approximate relation for 
 the high momentum distribution of protons and neutrons in the nucleus A\cite{proa2}:
\vspace{-0.2cm}
\begin{equation}
 n^{A}_{p/n}({p}) = {1\over 2 x_{p/n}} a_2(A,y)\cdot n_d(p)
 \label{highn}
 \vspace{-0.2cm}
 \end{equation}
 where $x_{p/n} = {Z\over A}/{A-Z\over A}$ and $y = |1-2x_p|$.  According to this relation  one expects {\em more}
 energetic protons than neutrons in  nuclei with an excess of neutrons ($x_n> x_p$).
 In the recent study\cite{proa2} the analysis of  the  existing  data demonstrated that 
 $a_2(A,y)$ is proportional to A and  decreases with an  increase of nuclear asymmetry ($y\rightarrow 1$).
   
Summarizing,  the recent SRC studies indicate that,  (a)  $a_2(A,y)$ is proportional to A and (b)  for large A 
due to the excess of neutrons more protons occupy the high momentum tail of the momentum distribution than 
neutrons.

 \begin{table}[t]
\begin{tabular}{llllll}
    \tablehead{1}{l}{b}{A}
  & \tablehead{1}{l}{b}{$P_p(\%)$}
  & \tablehead{1}{l}{b}{$P_n(\%)$}\
  &  \tablehead{1}{l}{b}{A}
  & \tablehead{1}{l}{b}{$P_p(\%)$}
  & \tablehead{1}{l}{b}{$P_n(\%)$}\\
 \hline
12  & 20 & 20 & 56   & 27 & 23\\
27  & 23 & 22 & 197 & 31 & 20 \\
\hline
\end{tabular}
\label{tab:a}
\end{table}

{\bf Correlation between EMC and SRC Effects:} One of the most intriguing  recent observations is the apparent correlation 
between the strength of the EMC effects (measured as $-dR_{EMC}/dx)$) and the strength of the SRCs 
(measured through $a_2$).  The initial observation\cite{Larry1} was that the correlation is purely linear,  however the most 
recent measurements indicate  on possibly of  non-linearity in  these correlations\cite{Larry2}.  To understand the reason for 
such correlation we explore two  possibilities; (a) it is the reflection of the fact that larger is $a_2$ smaller is the 
overall normalization of the mean-filed part of the momentum distribution and so is $R_{EMC}$ due to 
sizable mean-field contribution to the DIS cross section at x<1  and ({b}) if EMC effect is only due to the high momentum 
component of nuclear wave function then large $a_{2}$ will correspond to more medium modification therefore 
to smaller $R_{EMC}$.  

In the case of ({b}) in  calculations it is important to take into account Eq.(\ref{highn}) according to which 
protons and neutrons will have different amount of high momentum components  in asymmetric nuclei.  In the table 
we present the overall fractions of  high momentum ($>k_F$) protons and neutrons estimated according to Eq.(\ref{highn}).
This result indicates that for large A  the protons in average will be more energetic and virtual.
Since at $x>0.5$  proton DIS structure functions are larger than that of neutron  and if EMC effect is proportional 
to the nucleon virtuality, then we predict that the most of the EMC effect will be due to proton modification in the medium

\begin{figure}[ht]
  \includegraphics[height=.21\textheight]{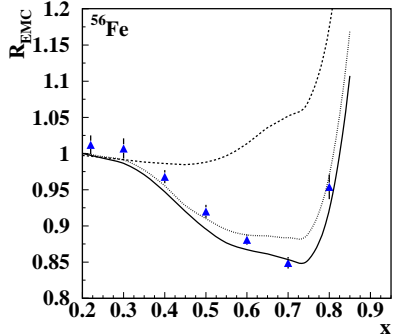}
  \caption{The x dependence of $R_{EMC}$ for $^{56}Fe$ nucleus.}
\end{figure}

{\bf The EMC Model:}
To be able to check the above conjectures we need to estimate $R_{EMC}$ 
within the model in which the effect is proportional to the virtuality of bound nucleon.  This model is developed based on 
the light cone~(LC) approximation of deep inelastic $eA$ scattering in which nuclear DIS structure function is expressed as\cite{FS8188,nestfun}:
\vspace{-0.2cm}
\begin{equation}
F_2^A(x,Q^2) = \sum\limits_{N=1}^{A} \int\limits_{x}^A{d\alpha\over\alpha}F^{bound}_{2N}({x\over\alpha},Q^2)n^A_N(\alpha),
\vspace{-0.2cm}
\label{lcd}
\end{equation}
where $\alpha$ is  (A times) the LC momentum fraction of the nucleus carried by the bound nucleon.  In the 
nuclear LC density matrix, $n^A_N(\alpha)$, the high momentum component is constructed according to 2N SRC model (Eq.\ref{highn}) 
which  contains the above mentioned asymmetry for protons and neutrons. The DIS structure function of 
the bound nucleon $F^{bound}_{2N}(x,Q^2)$ contains all the effects due to nuclear modification. 

For the nuclear modification we consider
the color screening model\cite{FS85} which is based on the observation that  the most significant EMC effect  
is observed at large  $x>0.5$  corresponding to  high momentum component of the quark distribution in the nucleon, in 
which three quarks are close together in point-like configurations ($PLC$). 
It is then assumed that the dominant contribution to  $F^{bound}_{2N}(x,Q^2)$  is given by $PLCs$ 
which, due to color screening,  interact weakly  with the other nucleons. 
As a result the  optimally bound configuration of nucleons will have suppressed contribution from the PLC component of nucleon wave function. 
This  suppression of $PLC$ in a bound nucleon is assumed to be the main source of the $EMC$ effect in inclusive $DIS$.  
 The suppression factor is calculated in perturbation series of the parameter: $\kappa = \left| {\langle U_A \rangle \over \Delta E_A}  \right|$, where   $\langle U_A \rangle$ is the average nuclear potential energy per nucleon  
 and $\Delta E_A \approx M^* - M \sim 0.6 \div 1 ~ GeV$ is the typical energy for nucleon excitations within the nucleus.
 The $PLC$ suppression can be represented  by  a multiplicative  factor $\delta_A(k^2)$   to  $F_2^N(x, Q^2)$ 
 that enters  in Eq.\ref{lcd}\cite{FS85}:
 \vspace{-0.2cm}
 \begin{equation}
        \delta_A(p^2) = {1\over (1+\kappa)^2} = {1 \over [1 + (p^2 / M + 
        2\epsilon_A) / \Delta E_A]^2}, \nonumber \\ 
\vspace{-0.2cm}
        \label{eq:gamma}
\end{equation}
where $p$ is the momentum of the bound nucleon in the light cone.

Using the above estimate of the suppression factor  we present in Fig.1  the comparison of our calculations\cite{emcsrc}  of $R_{EMC}$ for $^{56}Fe$ 
with the  data\cite{EMC5}.  As the comparisons show the prediction of LC dynamics (dashed curve),  in which no 
medium modifications are accounted for, 
grossly overestimates $R_{EMC}$.  The dash-doted curve accounts for the medium modification effects only due to the 
nuclear mean field and the solid line
includes, in addition, the modification of the nucleon DIS structure function in  SRC.  As the figure shows the modification in SRC becomes increasingly important at
$x>0.5$.  It is worth noting that the  present calculations are preliminary and does not account for the effects due to the 
Coulomb field\cite{Frankfurt:2010cb}  discussed earlier.  The latter effect as expected will shift the EMC strength towards 
higher x$(>0.5)$ thereby enhancing the role of the nucleon modifications in  SRC.

Finally we used our model\cite{emcsrc} to calculate the correlation between  $-dR_{EMC}/dx$ and $a_2$.  Our preliminary calculation (Fig.2)  
describes the correlation observed in Ref.\cite{Larry2} surprisingly well.  Both,  mean field depletion due to SRC and large medium 
modification in the SRC contribute to the calculated correlation. 
We predict nonlinear correlation and the main reason of this 
is the enhancement of the proton contribution in the EMC effect due to the increase of their average momenta in  large $A$ asymmetric nuclei (as it was discussed above).
Note that the Coulomb effects will change the current result slightly,  since our main effect is due to SRC  which is dominated 
at large $x>0.5$.

\begin{figure}[ht]
  \includegraphics[height=.21\textheight]{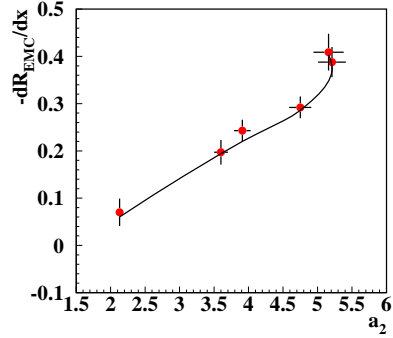}
  \caption{Correlation between the EMC slope and SRC strength.}
\end{figure}

{\bf Conclusion and Outlook:}
We present the first attempt to quantify the observed correlation between the strengths of the EMC effect and SRCs.  Our calculations show that two main 
factors contribute to  this correlation.  One,  with the increase of $a_2$ the normalization of the mean-field part of the nuclear wave function is decreasing 
which results to the   depletion of $R_{EMC}$. Second,  at $x>0.5$ the modification of the nucleons in the SRC plays increasingly important role and 
in all  models in which the EMC effect is proportional to nucleon virtuality the large effect of the EMC will be correlated to the large value of $a_2(A)$.  Finally 
we predict large EMC effect due to the increased proton contribution to $R_{EMC}$ for large A asymmetric nuclei.  This predication allows us to describe the 
non-linearity of the  EMC-SRC correlation  at large A.  

The new prediction of the increased role of the protons in the EMC effect can, in principle,  be  checked in deep inelastic semi-inclusive $(A(e,e'N)X$ reactions 
in which the spectator nucleon is detected in the backward hemisphere of the reaction which minimizes the final state interaction effects\cite{tagged,polext,Cosyn}.

{\bf Acknowledgments:}This work is supported by U.S. Department of Energy grant under contract DE-FG02-01ER41172.

\bibliographystyle{aipproc}   

\end{document}